\documentclass{article}

\usepackage{graphicx}
\usepackage{dcolumn}
\usepackage{bm}
\usepackage{graphics}
\usepackage{graphicx}
\usepackage{amsmath,amssymb}
\usepackage{color}

\title{Analyzing Chaos in Higher Order Disordered Quartic-Sextic Klein-Gordon Lattices Using $q$-Statistics}
\author{Chris G. Antonopoulos\\Department of Mathematical Sciences,\\University of Essex, UK\\\\Charalampos Skokos\\Department of Mathematics and Applied Mathematics,\\University of Cape Town, South Africa\\\\Tassos Bountis\\Department of Mathematics, School of Science and Technology,\\Nazarbayev University, Kazakhstan\\\\Sergej Flach\\Center for Theoretical Physics of Complex Systems,\\Institute for Basic Science, South Korea}

\date{\today}
\begin{document}

\maketitle

\begin{abstract}
In the study of subdiffusive wave-packet spreading in disordered Klein-Gordon (KG) nonlinear lattices, a central open question is whether the motion continues to be chaotic despite decreasing densities, or tends to become quasi-periodic as nonlinear terms become negligible. In a recent study of such KG particle chains with quartic (4th order) anharmonicity in the on-site potential
it was shown that $q-$Gaussian probability distribution functions of sums of position observables with $q > 1$  always approach pure Gaussians ($q=1$) in the long time limit and hence the motion of the full system is ultimately ``strongly chaotic''. In the present paper, we show that these results continue to hold even when a sextic (6th order) term is gradually added to the potential and ultimately prevails over the 4th order anharmonicity, despite expectations that the dynamics is more ``regular'', at least in the regime of small oscillations. Analyzing this system in the subdiffusive energy domain using $q$-statistics, we demonstrate that groups of oscillators centered around the initially excited one (as well as the full chain) possess strongly chaotic dynamics and are thus far from any quasi-periodic torus, for times as long as $t=10^9$.
\end{abstract}

\section{Introduction}\label{sec_intro}

Anderson localization \cite{Ander}, i.e.~the absence of wave diffusion in disordered media, is a phenomenon affecting many physical processes, such as for example the conductivity of materials, the behavior of granular matter, the dynamics of Bose-Einstein condensates, etc. The effect of nonlinearity on the localization properties of wave packets in disordered systems has attracted the attention of many researchers, both experimentally
\cite{SBFS07,LAPSMCS08,REFFFZMMI08}
 and theoretically \cite{pikovsky_destruction_2008,kopidakis_absence_2008,Flach2009,Skokos2009,veksler_spreading_2009,GMS09,SF10,Flach_spreading_2010, Laptyeva_crossover_2010,VKF10,MP10,MAP11,PF11,BLSKF11,BLGKSF11,B11,BouSkobook2012,MP12,LBF12,MP13,SGF13,Antonopoulosetal2014,LIF14,ILF14,F16}. Recent studies of nonlinear disordered variants of two typical one-dimensional Hamiltonian lattice models, namely the Klein-Gordon (KG) oscillator chain and the discrete nonlinear Schr\"{o}dinger equation, revealed the statistical characteristics of energy spreading and showed that nonlinearity destroys localization \cite{Flach2009,Skokos2009,SF10,Laptyeva_crossover_2010,BLSKF11,BLGKSF11}. In these papers, the existence of different dynamical behaviors in different energy density regimes was established, their particular dynamical characteristics were determined and their appearance was theoretically explained.

However, important questions regarding the asymptotic behavior of wave-packet spreading, and the persistence of chaos in such systems,  still remain unanswered. Some researchers have conjectured \cite{Johansson_kam_2010,A11} that wave packets will eventually approach torus-like structures in phase space, exhibiting at the same time less chaotic behavior which eventually leads to the halt of energy spreading in the chain. Although most numerical investigations on nonlinear disordered lattices show that wave packets continue spreading chaotically, at least up to times accessible to computer simulations, some numerical indications of a possible slowing down of spreading for particular models have been reported in the literature \cite{PF11,MP13}. The diversity
of the models studied so far includes systems with different numbers $N_{IoM}$ of integrals of motion or conserved quantities, and the degree of anharmonicity $\sigma$ which is related to a corresponding $n$-body interaction  or
equally to the number of interacting normal modes mediated by the anharmonicity. $N_{IoM}=1$ for the mentioned
case of KG lattices, while $N_{IoM}=2$ for the DNLS case. In most studied cases $\sigma=2$ which corresponds to quartic anharmonicity and two-body interactions. Here we will also study $\sigma=4$ which corresponds to
sextic anharmonicity and three-body interactions.

To investigate the potential asymptotic approach to regular (or irregular) dynamics, the properties of the motion in the subdiffusive regime of the KG model with $\sigma=2$ were recently studied \cite{SGF13,Antonopoulosetal2014}. In \cite{SGF13}, the computation of the maximum Lyapunov exponent (MLE) showed that although chaotic dynamics slows down as expected from a subdiffusive process, it does not cross-over to a regime of regular behavior. In \cite{Antonopoulosetal2014} the dynamics of a disordered quartic (termed KG4) lattice was studied using $q$-statistics \cite{Tsallisbook2009} by analyzing probability distribution functions (pdfs) of position observables describing the evolution of wave packets initiated by exciting the central lattice particle. In that work it was shown that the overall motion displays strongly chaotic behavior for long times, again with no signs of the dynamics relaxing on quasi-periodic tori.

In this paper, we apply the above methodology followed in \cite{Antonopoulosetal2014,SGF13} to a KG model which includes a gradually increasing sextic anharmonicity which is assumed to show an asymptotic slowing down of the wave packet spreading - if existing - at earlier times, since the impact of this anharmonicity is weaker than the quartic one, as long as small densities are considered.
In line with what we discovered in the KG4 model \cite{Antonopoulosetal2014}, both the full lattice, as well as groups of particles around the initially excited one at the center remain strongly chaotic and show no signs of approaching quasi-periodicity, even after very long integration times (at least up to $t=10^9$, with  time scales
dictated by the linear equations being of order one). On the other hand, individual particles far from the center, after interacting with the wave-packet behave at first weakly chaotically, but later also tend to strong chaos for times as long as $t=10^9$. We find that the closer the particles are to the center of the lattice, the more strongly chaotic their behavior is (with pdfs closer to the Gaussian $q=1$ case), and that the dynamics of the whole lattice is always strongly chaotic, with pdfs obeying Boltzmann-Gibbs thermostatistics. Thus, all our results indicate that the wave packet spreading is a truly chaotic process which does not show any tendency to  become more regular.
Our results are also supported by the computation of the time dependence of the largest Lyapunov exponent
which again show no cross-over into a regime of regular behavior, similar to the KG4 case \cite{SGF13}.

The paper is structured as follows: In Sec.~\ref{model_methods} we present the KG model and outline the
used statistical methods in the spirit of the Central Limit Theorem (CLT). In Sec.~\ref{sec_results1} we examine the mixed case of both quartic and sextic anharmonicities in the potential and consider the statistical properties of the dynamics as the sextic terms become increasingly more important. In Sec.~\ref{sec_results2}, we focus on the purely sextic anharmonicity model and describe the results obtained when we excite only the central particle of a $500$ particle chain and monitor the chaotic evolution of individual particles, groups of particles about the central one, as well as the whole system. Our conclusions and a discussion follow in Sec.~\ref{section_conclusions}.

\section{Model and numerical methods}\label{model_methods}

The Hamiltonian of the disordered one-dimensional KG lattice studied in \cite{SF10} is
\begin{equation}
H= \sum_{l=1}^{N} \frac{p_{l}^2}{2} + \frac{\tilde{\epsilon}_{l}}{2}
x_{l}^2 + \frac{|x_{l}|^{\sigma+2}}{\sigma+2} +\frac{1}{2W}(x_{l+1}-x_l)^2,
\label{RQKG}
\end{equation}
where $l$ is the lattice site index, $x_l$ and $p_l$ are respectively the generalized canonically conjugated coordinates and momenta (with $x_{N+1}=0$), $\sigma$ measures the degree of anharmonicity and $W=4$ controls the nearest neighbour
interaction strength and thus the effective strength of disorder. Disorder enters through the onsite harmonic squared  frequencies $\tilde{\epsilon}_{l}$ which are random uncorrelated numbers chosen uniformly from the interval $\left[\frac{1}{2},\frac{3}{2}\right]$. The total energy $E \equiv H \geq 0$ of the system serves as a control parameter of the nonlinearity for fixed disorder strength $W$. The case $\sigma=2$ corresponds to the typical quartic disordered KG4 model. Wave packet spreading in the KG4 case was studied in several papers \cite{Flach2009,Skokos2009,Laptyeva_crossover_2010,BLSKF11,BLGKSF11,SGF13,Antonopoulosetal2014}.
The equations of motion follow as $\dot{x}_l = \partial H / \partial p_l$ and $\dot{p}_l = -\partial H / \partial x_l$.

The dynamics of wave packet spreading in the Hamiltonian (\ref{RQKG}) was analyzed in detail in \cite{SF10} for
several anharmonicity values $\sigma$, following the evolution of the normalized energy density
\begin{equation}\label{eq:z}
    z_{l}= \frac{E_{l}}{\sum_{i=1}^{N} E_{i}}
\end{equation}
of the site-energies $E_l=\frac{p_{l}^2}{2} + \frac{\tilde{\epsilon}_{l}}{2}
x_{l}^2 + \frac{|x_{l}|^{\sigma+2}}{\sigma+2} +\frac{1}{4W}\left[ (x_{l+1}-x_l)^2 +(x_{l-1}-x_l)^2\right]$.
The main goal was to monitor the evolution of the second moment $m_2$
\begin{equation}\label{second_moment}
	m_2= \sum_{l=1}^{N} (l-\bar{l})^2 z_{l}\;,\;\;\;\; \bar{{l}} = \sum_{l=1}^{N} l z_{l}
\end{equation}
which quantifies the wave packet
degree of spreading, and the participation number
\begin{equation}\label{PN}
P=\frac{1}{\sum_{l=1}^{N} z_l^2},
\end{equation}
which measures the number of the most strongly excited sites in the system.
In \cite{SF10} it was found that the wave packet spreads incoherently and subdiffusively
with $m_2\propto t^\alpha$ and $P\propto t^{\alpha/2}$ with the exponent
\begin{equation} \label{exponent}
\alpha=\frac{1}{\sigma+1}
\end{equation}
up to the largest computed times, indicating no loss of incoherent (chaotic) dynamics and no cross-over to
coherent (regular) dynamics.

In the present paper we consider a hybrid model (termed KG46 here) which interpolates between the quartic $\sigma=2$ and
sextic $\sigma=4$ cases by including sextic terms  in \eqref{RQKG}. We expect that sextic terms are generating
weaker nonlinear terms in the equations of motion, and could amplify a crossover to regular dynamics, if present.
In all our models we follow the evolution of single-site excitations of the central particle in the subdiffusive regime defined in \cite{SF10} (see Fig.~1 in \cite{SF10}), by considering $N=500$ sites and setting the total energy of our lattice to some constant energy $E$. In particular, for a given set of $\tilde{\epsilon}$ values, we
choose $x_l=0$ and $p_l=\sqrt{2E}\delta_{l,N/2}$, where $\delta_{i,j}$ is the Kronecker delta, thus exciting precisely one oscillator in the center of the system
which has $N$ sites.

To study the resulting trajectories, we integrate numerically the equations of motion of Hamiltonian \eqref{RQKG} by the 4th order Yoshida's symplectic integrator \cite{Yoshida1990}. In our simulations, we set the integration time step to $\tau=0.05$, which typically keeps the relative energy error at about $10^{-6}$. Furthermore, to obtain reliable statistical results that are independent of the particular realizations, we consider an ensemble of 64 disorder realizations, i.e.~64 random sequences of $\tilde{\epsilon_{l}}$ values in \eqref{RQKG}. Apart from the computation of $m_2$ and $P$, we also evaluate the MLE $\lambda_1$. For this purpose, we use the same symplectic integration scheme for the integration of the variational equations of system \eqref{RQKG} according to the {\it tangent map method} \cite{Skokosetal2010,Gerlachetal2011,GES12}. The variational equations govern the evolution of small deviation vectors from the studied trajectory and are used for the evaluation of the MLE according to the so-called standard method \cite{Benettin1980a,Benettin1980b,S10}.

We use the solutions of the equations of motion of Hamiltonian \eqref{RQKG} to construct pdfs of suitably rescaled sums of $M$ values of a generic observable $\eta(t_i),\;i=1,\ldots,M$, which depends linearly on the position coordinates of the solution. Viewing these as independent and identically distributed (iid) random variables (in the limit of $M\rightarrow\infty$), we evaluate their sum
\begin{equation}\label{sums_CLT}
S_M^{(j)}=\sum_{i=1}^M\eta(t_i)^{(j)},\;j=1,\ldots,N_{\mbox{ic}},
\end{equation}
for $N_{\mbox{ic}}$ initial conditions and study the statistics of quantities \eqref{sums_CLT} centered about their mean value $\langle S_M^{(j)}\rangle$ and rescaled by their standard deviation
\begin{equation}
s_M^{(j)}\equiv\frac{S_M^{(j)}-\langle S_M^{(j)}\rangle}{\sigma_M},
\end{equation}
where $\sigma_M^2=\langle S_M^{(j)2}\rangle -\langle S_M^{(j)}\rangle^2$. Plotting the normalized histogram of the probabilities $\mathtt{P}(s_M^{(j)})$ as a function of $s_M^{(j)}$, we then compare the resulting numerically computed pdfs with a $q$-Gaussian of the form
\begin{equation}\label{q_gaussian}
\mathtt{P}(s_M^{(j)})=a\exp_q({-\beta s_M^{(j)2}})\equiv a\biggl(1-(1-q)\beta s_M^{(j)2}\biggr)^{\frac{1}{1-q}},
\end{equation}
where $q$ is the so-called entropic index, $\beta$ is an arbitrary parameter and $a$ a normalization constant. Equation \eqref{q_gaussian} is a generalization of the well-known Gaussian pdf, since in the limit $q\rightarrow 1$ we have $\exp_q(-\beta x^2)\rightarrow\exp(-\beta x^2)$. Moreover, it can be shown that the $q$-Gaussian distribution is normalized when
\begin{equation}\label{beta-$q$-Gaussian}
\beta=a\sqrt{\pi}\frac{\Gamma\Bigl(\frac{3-q}{2(q-1)}\Bigr)}{(q-1)^{\frac{1}{2}}\Gamma\Bigl(\frac{1}{q-1}\Bigr)},
\end{equation}
where $\Gamma$ is the Euler $\Gamma$ function. Clearly, equation \eqref{beta-$q$-Gaussian} shows that the allowed values of $q$ are $1<q<3$ for the normalization to be possible.

The index $q$ appearing in \eqref{q_gaussian} is connected with the Tsallis entropy \cite{Tsallisbook2009}. Systems characterized by the Tsallis entropy are said to lie at the ``edge of chaos'' and are significantly different from Boltzmann-Gibbs systems, in the sense that their entropy is non-additive and generally non-extensive \cite{Tsallisbook2009}.

Let us now consider the values of one (or a linear combination) of coordinates $x$ of the solution of system \eqref{RQKG} at discrete times $t_i,\;i=1,\ldots, M$, as realizations of $N_{\mbox{ic}}$ random variables $X^{(j)}(t_i)$, $j=1,\ldots,N_{\mbox{ic}}$ and study in detail their statistics. Typically, the results presented in this work are obtained by setting $N_{\mbox{ic}}=10^6$ and $M=10^3$. According to the CLT \cite{Rice1995}, if these variables are random, their sum-distributions yield a Gaussian pdf, whose mean and variance are those of the $X^{(j)}$'s. This happens for many dynamical systems in strongly chaotic regimes, where correlations decay exponentially and the system obeys Boltzmann-Gibbs statistics. However, in weakly chaotic regimes, these pdfs {\it do not} rapidly converge to a Gaussian, but instead are approximated for long times, by a $q$-Gaussian distribution \eqref{q_gaussian} with $1\leq q<3$.

\section{Results for the hybrid quartic-sextic (KG46) case}\label{sec_results1}

Let us consider a Hamiltonian that includes quartic as well as sextic order terms in the on-site potential, i.e.
\begin{equation}
H= \sum_{l=1}^{N} \frac{p_{l}^2}{2} + \frac{\tilde{\epsilon}_{l}}{2}
x_{l}^2 + A\frac{x_{l}^{4}}{4} + (1-A)\frac{x_{l}^{6}}{6} +\frac{1}{2W}(x_{l+1}-x_l)^2,
\label{RQKG46}
\end{equation}
where $0\leq A\leq 1$. Note that the introduction of the parameter $A$ in (\ref{RQKG46}) allows us to investigate a family of Hamiltonians that ranges from the KG4 ($A=1$) to the KG6 ($A=0$) system.

\begin{figure}[]
\begin{center}
\includegraphics[width=0.4\textwidth]{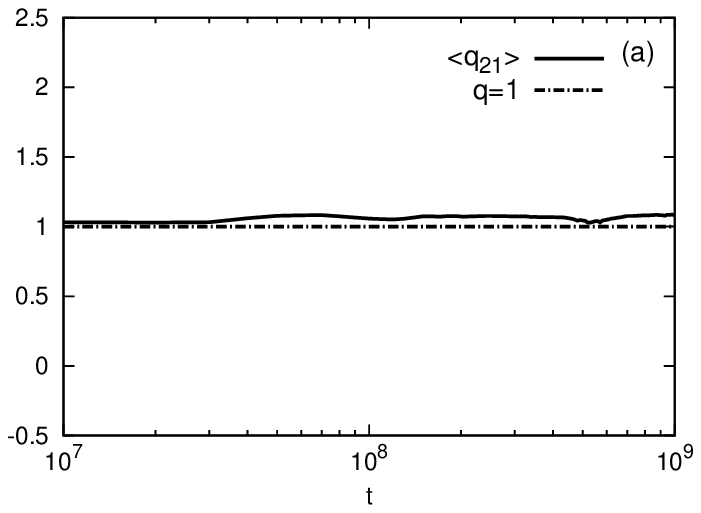}
\includegraphics[width=0.4\textwidth]{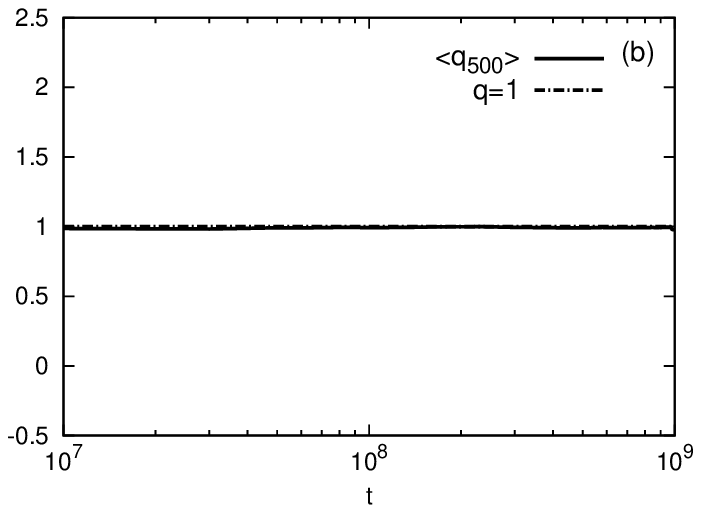}
\includegraphics[width=0.4\textwidth]{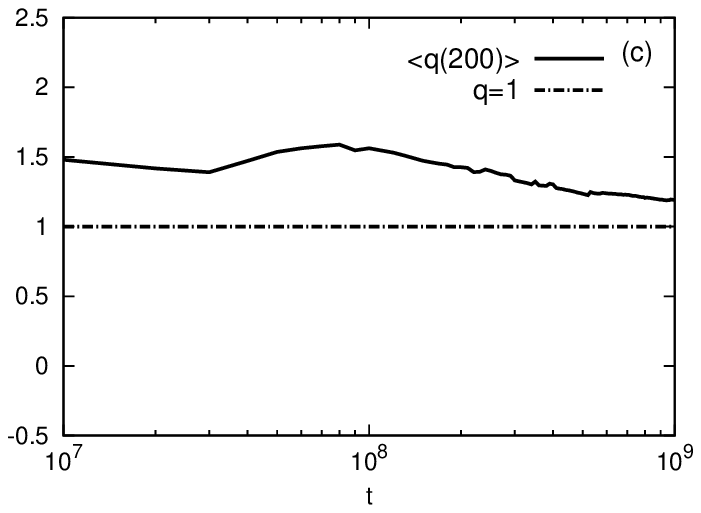}
\includegraphics[width=0.4\textwidth]{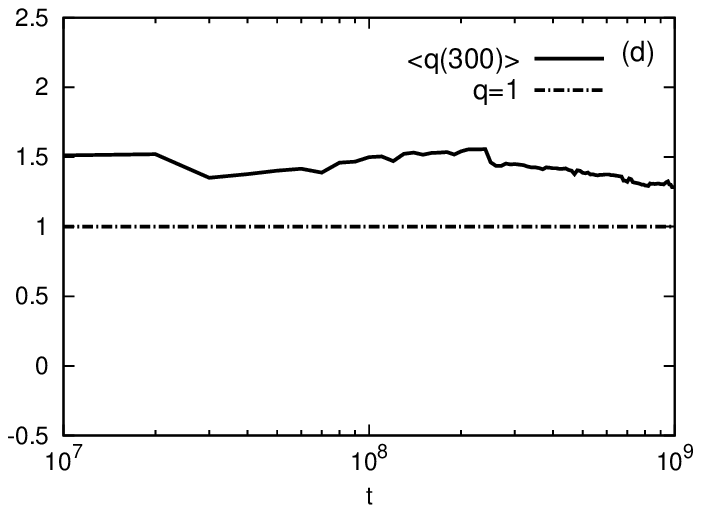}

\includegraphics[width=0.4\textwidth]{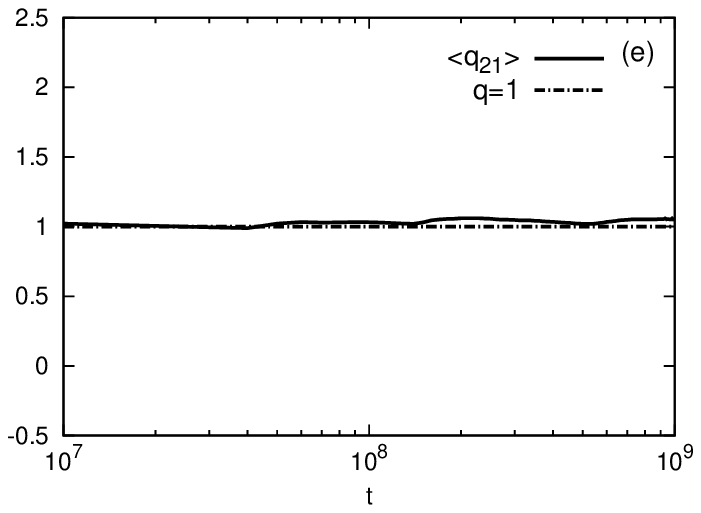}
\includegraphics[width=0.4\textwidth]{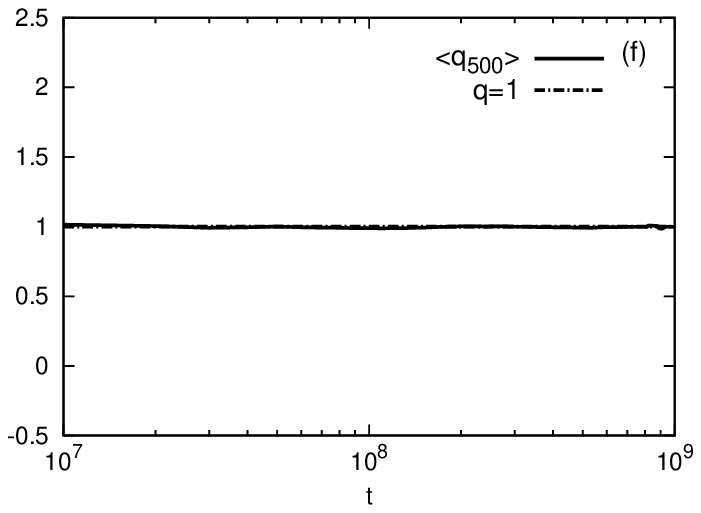}
\includegraphics[width=0.4\textwidth]{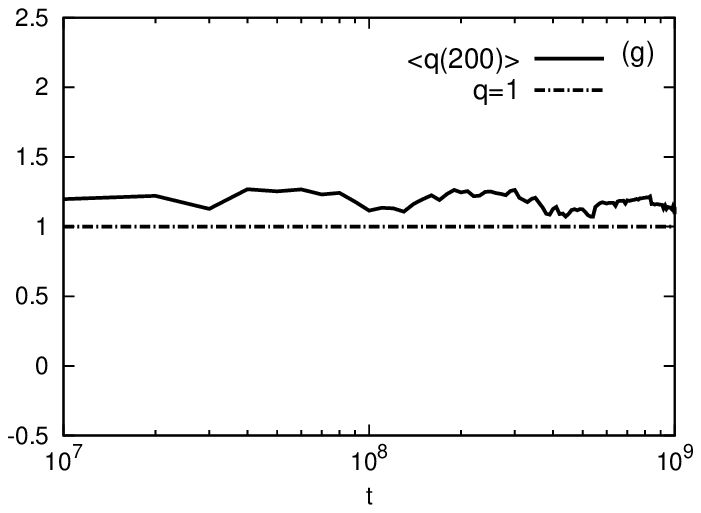}
\includegraphics[width=0.4\textwidth]{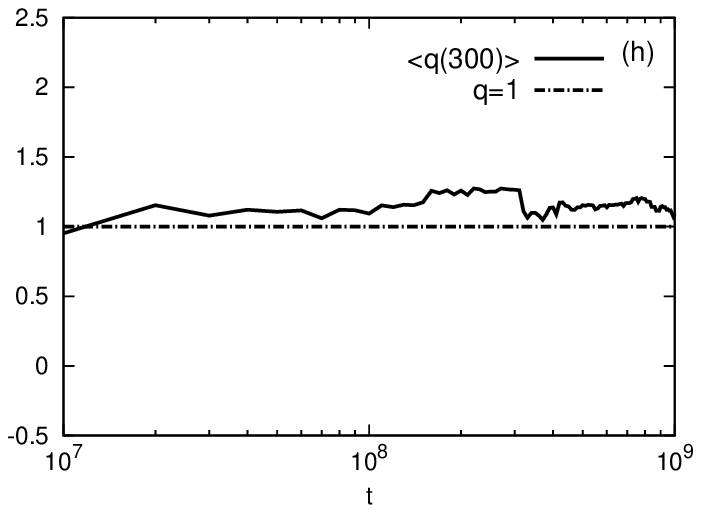}
\end{center}
\caption{Entropic indices averaged over 64 random disordered realizations for the hybrid quartic-sextic KG46 chain of $N=500$ particles with $A=0.5$, as well as the purely sextic KG6 chain with $A=0$. We first present indicative results for the mixed case $A=0.5$: (a) $\langle q_{21} \rangle$, 21 particles around the central particle and (b) $\langle q_{500} \rangle$, the whole chain, as well as individual particles in (c) $\langle q(200) \rangle$ and (d) $\langle q(300) \rangle$ for the 200th and 300th particles respectively. We also show entirely analogous results for the case $A=0$ in (e)-(h). All simulations correspond to single site excitations of the central particle $i=250$. All pdfs considered for the evaluation of the entropic indices are computed for 100 time windows in $[0,t_f]$, where each time window starts always from $t=0$.}
\label{fig_A05}
\end{figure}

Following \cite{Antonopoulosetal2014}, we have used energy values $E=0.4$ and $E=0.6$ in (\ref{RQKG46}), which corresponds to the subdiffusive regime of the KG4 system. Although we studied the behavior of system (\ref{RQKG46}) for various values of parameter $A$, we present in Fig.~\ref{fig_A05} illustrative results for $A=0.5$
only. Figure \ref{fig_A05} shows the time evolution of the entropic index averaged over 64 realizations for four indicative cases: $ \langle q_{21} \rangle$ corresponding to 21 particles about the center, $\langle q_{500} \rangle$ for the full chain, and $\langle q(200) \rangle$ and $\langle q(300) \rangle$ for the 200th and 300th particles. Here $\langle ... \rangle$ indicates averaging of the quantity over all 64 disorder realizations. We first plot these indices in Fig.~\ref{fig_A05}(a)-(d) for the hybrid quartic-sextic KG46 case $A=0.5$ and then repeat the calculation for the same indices in the purely sextic KG6 case $A=0$ in Fig.~\ref{fig_A05}(e)-(h). The only difference is that particle groups have indices that rapidly converge to $q=1$, while individual particles behave weakly chaotically with indices significantly higher than unity, but also show a tendency to approach $q=1$ for very long times. From these results we conclude that the dynamics of the chain is strongly chaotic as the $q$-values are close to 1, especially when large parts of the lattice are considered, just as was observed in the purely quartic KG4 case studied in \cite{Antonopoulosetal2014}.

Now, for any non-zero values of $A$ it is expected that the asymptotic behavior of the dynamics will be controlled by the quartic terms of the Hamiltonian (or equivalently by the cubic terms in the corresponding equations of motion),
since the contribution of the sixth order part will eventually become negligible as the wave packet spreads and the values of $x_l$ decrease in magnitude. This is indeed what we observe for $A=0.5$ in Fig.~\ref{fig_A05}(a),(b) since the overall dynamics is similar to the behavior of the KG4 model studied in \cite{Antonopoulosetal2014}. This turns out to be more generally true as the results for several other non-zero values of $A$ (not presented here) are qualitatively similar to the ones shown in  Fig.~\ref{fig_A05}(a)-(d). For this reason, we focus
our attention
in the next section on the $A=0$ case.

\section{Results for the purely sextic KG6 case}\label{sec_results2}

\begin{figure}[]
\begin{center}
\includegraphics[width=0.49\textwidth]{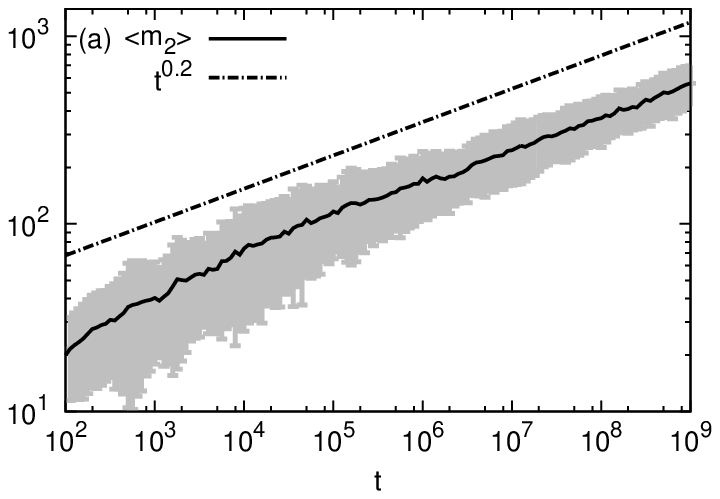}
\includegraphics[width=0.49\textwidth]{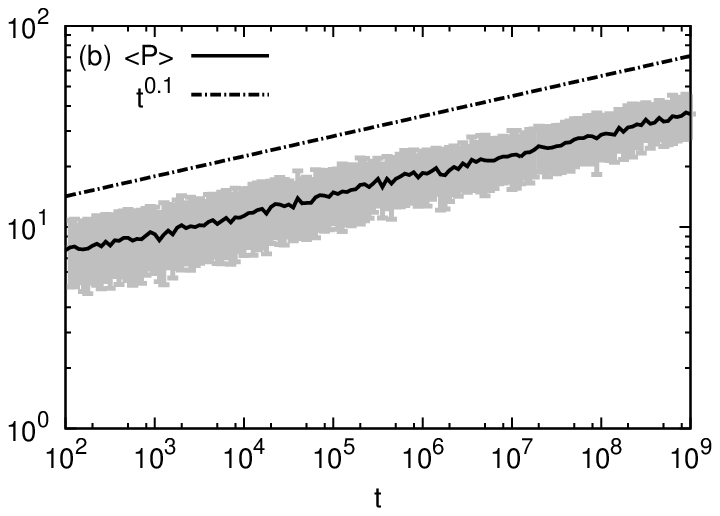}
\end{center}
\caption{The KG6 chain $(A=0)$ averaged (over 64 disordered realizations). (a) the second moment $\langle m_2\rangle$ versus time, and (b) the participation number $\langle P\rangle$ versus time. In both plots, we show in grey one standard deviation around the averaged curves, while the dashed-dotted line in (a) corresponds to the power law $t^{0.2}$ and the dashed-dotted line in (b) corresponds to $t^{0.1}$. The power law in panel (b) is half of that in panel (a), and both are in agreement with \cite{SF10} and Eq. \eqref{exponent}.}
\label{fig_subdiffusive_behavior1}
\end{figure}

\begin{figure}[]
\begin{center}
\includegraphics[width=0.49\textwidth]{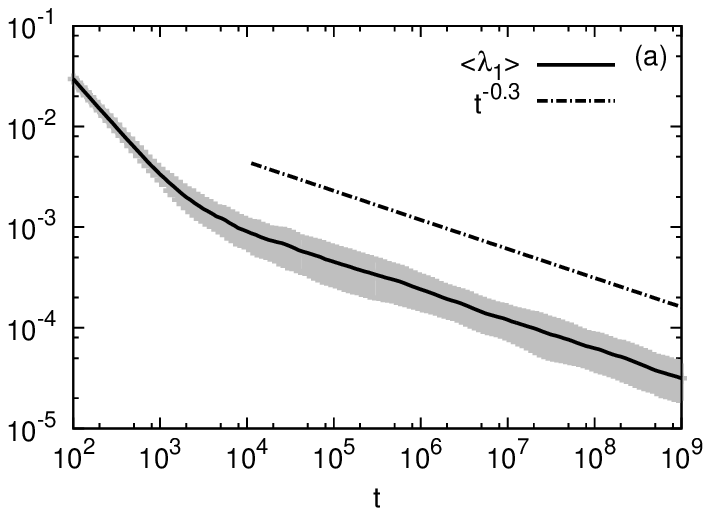}
\includegraphics[width=0.49\textwidth]{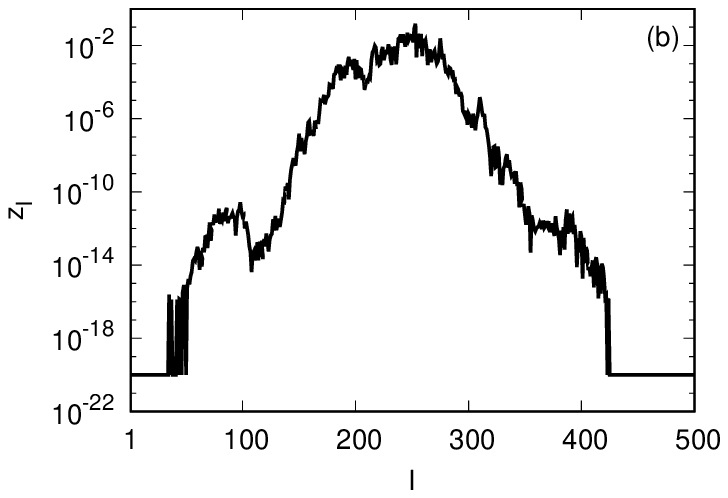}
\end{center}
\caption{Subdiffusive motion of single-site excitations in the KG6 chain $(A=0)$ averaged (over 64 disordered realizations). (a) Here the maximum Lyapunov exponent MLE $\langle\lambda_1\rangle$ is shown, while (b) depicts the normalized energy $z_l$ of all particles $l=1,\ldots,500$ of one of the 64 realizations, at $t=10^7$. In panel (a) we plot in grey one standard deviation around the averaged curve, while the dashed-dotted line corresponds to the power law $t^{-0.3}$.}
\label{fig_subdiffusive_behavior2}
\end{figure}

We set $A=0$ in (\ref{RQKG46}), and present in Figs.~\ref{fig_subdiffusive_behavior1} and \ref{fig_subdiffusive_behavior2} the subdiffusive indicators of single-site excitations of the central particle averaged over 64 disorder realizations. First, in Fig.~\ref{fig_subdiffusive_behavior1}(a),(b) we show the time dependence of the second moment $\langle m_2\rangle$ and participation number $\langle P\rangle$ respectively, shading by grey one standard deviation away from the mean. Here $\langle ... \rangle$ indicates averaging of the logarithm of the quantity over all 64 disorder realizations. In Fig.~\ref{fig_subdiffusive_behavior2}(a) we show the MLE $\langle\lambda_1\rangle$, while in Fig.~\ref{fig_subdiffusive_behavior2}(b) we depict the normalized energy $z_l$ of all particles $l=1,\ldots,500$ for one of the 64 realizations, at $t=10^7$. As we see in Fig.~\ref{fig_subdiffusive_behavior2}(a), $\lambda_1$ first decays as $1/t$ during an initial transient phase, and after that continues to decay with a much weaker exponent as $1/t^{0.3}$,  which indicates that the motion remains chaotic in time (since regular motion would correspond to $1/t$ \cite{S10}). This behavior is similar to the one observed in \cite{SGF13} for the KG4 model, with a slightly different decay exponent $1/t^{0.25}$. The results in Fig.~\ref{fig_subdiffusive_behavior1}(b) are in good agreement with the prediction of Eq.~\eqref{exponent}. In Fig.~ \ref{fig_subdiffusive_behavior2}(b), on the other hand, it is evident that the particles at both edges of the chain are not excited at this time. We have also been careful to check that the normalized energies $z_l$ for particles close to the edges are always smaller than $10^{-10}$, at least up to $t=10^9$, to guarantee that the wave packet did not reach the edges of the chain during the integration times.

Let us now turn to the statistical properties of the KG6 lattice as the initial excitation of the central particle diffuses to its neighboring sites. In Fig.~\ref{fig_A05}(e),(f) we present for $A=0$ the time evolution of the same two entropic indices as in Fig.~\ref{fig_A05}(a),(b) for $A=0.5$, implying again that any collection of particles around the central one, including the full lattice, show strongly chaotic behavior as  their averaged entropic indices are practically equal to one. On the other hand, Fig.~\ref{fig_A05}(g),(h) show that single particle statistics is analogous to what one finds for $A=0.5$ in Fig.~\ref{fig_A05}(c),(d). These results suggest that the wave packet spreading dynamics in the subdiffusive regime of the KG6 system also does not approach a KAM regime of invariant tori. This is in agreement with \cite{Antonopoulosetal2014,SF10,SGF13}, where similar behaviors were found for KG4 systems, and with Sec.~\ref{sec_results1} where the hybrid KG46 models were considered, and thus supports the conclusions that the overall motion in these systems never approaches a KAM regime of quasi-periodic motion.

\begin{figure}
\begin{center}
\includegraphics[width=0.49\textwidth]{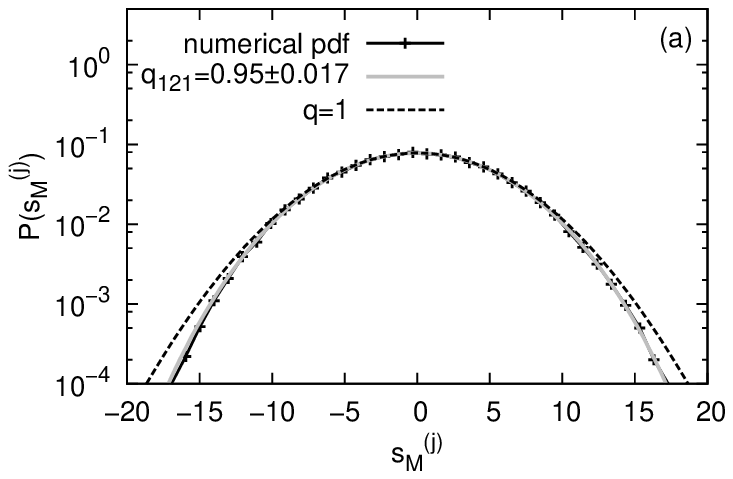}
\includegraphics[width=0.49\textwidth]{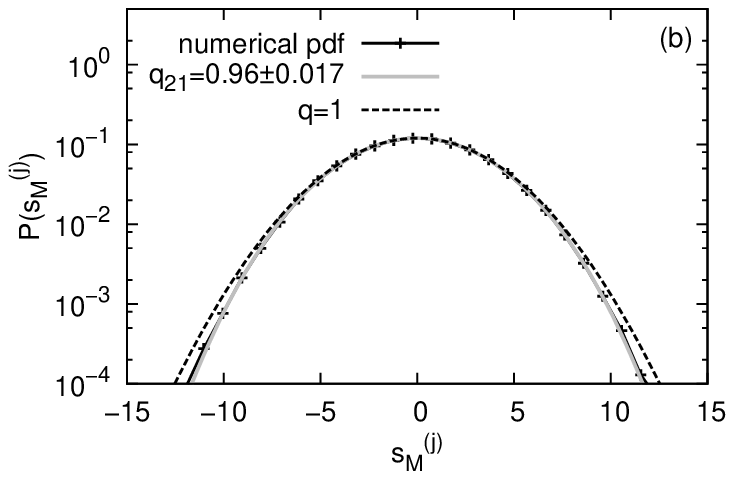}
\end{center}
\caption{Numerical pdfs (solid black + curve) for a group of 121 and 21 particles symmetrically located around the central particle for one disorder realization (panel (a)) of the KG6 model ($A=0$, 121 particles) and (panel (b)) for the hybrid KG46 model ($A=0.5$, 21 particles). The numerical pdf in panel (a) is computed for data obtained in the time interval $[9\cdot10^{6},10^7]$ and the pdf in panel (b) for data in $[0,9.9\cdot10^8]$. Both are approximated by $q$-Gaussian distributions of similar q-entropic indices (solid grey curves). For comparison, we also plot the $q=1$  Gaussian distribution (dashed curve). Note the similarity of the results in both panels.}
\label{fig_complex_statistics}
\end{figure}

As an illustrative example of how the corresponding pdfs look like, we present in Fig.~\ref{fig_complex_statistics}(a), for $A=0$, a pdf with $q$ slightly smaller than 1. This distribution refers to a group of 121 particles symmetrically located around the particle $i=250$ for one disorder realization of the KG6 model with $E=0.6$. The pdf is constructed from values of the observable $\eta_{121}=\sum_{i=190}^{310}x_{i}$  for $t\mbox{ in }[9\cdot10^{6},10^7]$. The analysis yields $q_{121}=0.95\pm0.017$, which is very close to the $q=1$ case of Boltzmann-Gibbs strong chaos. Entirely analogous results are obtained for the hybrid KG chain with $A=0.5$ in Fig.~\ref{fig_complex_statistics}(b) for 21 particles around the central one.

The significance of $q<1$ values is as yet unclear. Thus, we have performed an analysis with different summands for the calculation of the pdfs as a function of time and have found they exhibit smaller than 1 entropic indices for long time intervals. This suggests that $q<1$ may not be due to numerical errors as it is persistent for long enough times, but does not entirely exclude this possibility (e.g. finite time and size effects). Conversely, it may be due to a deeper reason related to a similar result reported recently in \cite{TirnakliTsallis2016}, where large systems of coupled logistic maps near their chaotic threshold were also found to exhibit pdfs with $q<1$ when subjected to a {\it different type of randomness}, namely that of additive noise in the equations of motion. Consequently, we believe that more work on this fascinating behaviour should be devoted to identify the underlying reason.

\section{Conclusions}\label{section_conclusions}

In this paper we have analyzed the chaotic properties of wave-packet spreading in KG disordered lattices with higher than 4th order nonlinearity in their on-site potential, using the methodology of $q$-statistics and non-extensive Statistical Mechanics \cite{Tsallisbook2009}. Our aim was to provide further evidence to strengthen the conclusions of earlier investigations \cite{Antonopoulosetal2014,SGF13} that wave packet spreading remains strongly chaotic for longer and longer integration times. Studying the entropic $q$ index of suitably chosen probability density functions, we showed numerically that the lattice in parts, but also as a whole, behaves strongly chaotically, with $q\rightarrow1$ as time increases to $t=10^9$. This suggests that the overall motion in these systems does not approach the quasi-periodic regime of invariant tori as conjectured by some authors, due to progressively smaller nonlinear effects at each lattice site as time goes to infinity.

To further establish this conclusion, we examined a class of on-site disordered potentials with quartic and/or
sextic anharmonicities, and analyzed wave packet dynamics generated by exciting the central particle and spreading subdiffusively over a 500-site one-dimensional chain. We found that all these KG systems remain strongly chaotic, in parts and as a whole, showing no sign of approaching quasi-periodic behavior, even after times as long as $t=10^9$. We also verified that individual particles close to the center of the lattice, after interacting with the wave, exhibit first weakly chaotic motion with $q$ values significantly larger than unity, but eventually also show a tendency towards strong chaos, just as was reported in \cite{Antonopoulosetal2014} for a KG4 model with purely quartic anharmonicity. Focusing on the strictly sextic KG6 model, which is expected to be closer to linear dynamics for small energy densities, we noted that here also strongly chaotic behavior prevails as time becomes very large.

In conclusion, the indisputable fact is that when we consider the full lattice or groups of particles around its center, the long-time dynamics is always strongly chaotic with entropic index $q=1$. Interestingly, in the purely
sextic KG6 case, as well in the hybrid KG46 lattice, there are particle groups that attain $q$ values slightly less than one (see Fig.~\ref{fig_complex_statistics})! This is reminiscent of a similar result in a recently published $q$-statistical analysis of coupled weakly chaotic logistic maps in the presence of noise \cite{TirnakliTsallis2016}. Of course, noise in \cite{TirnakliTsallis2016} enters additively in the evolution equations, while in our system disorder is {\it spatial} and present only in the parameters. Thus, the occurrence of $q<1$ in such models requires further study to shed more light on the different types of chaotic behavior occurring in these multi-dimensional Hamiltonian systems.

\section{Acknowledgements}\label{acknowledgements}
Ch. S. acknowledges support  by the National Research Foundation of South Africa (Incentive Funding for Rated Researchers, IFRR and Competitive Programme for Rated Researchers, CPRR).
This work was supported by the Institute for Basic Science of South Korea, project IBS-R024-D1.
We are grateful to Professor L. B. Drossos for many useful discussions and for helping us use the computational facilities of his HPCS-DL Lab at the Technological Educational Institute of Western Greece.





\providecommand{\noopsort}[1]{}\providecommand{\singleletter}[1]{#1}%

\end{document}